\documentclass[conference]{IEEEtran}
\IEEEoverridecommandlockouts
\usepackage{cite}
\usepackage{amsmath,amssymb,amsfonts}
\usepackage{algorithmic}
\usepackage{graphicx}
\usepackage{float}
\usepackage{textcomp}
\usepackage{url}
\usepackage{xcolor}
\def\BibTeX{{\rm B\kern-.05em{\sc i\kern-.025em b}\kern-.08em
    T\kern-.1667em\lower.7ex\hbox{E}\kern-.125emX}}
\begin{document}

\title{Assessing the Efficacy of IoT-based Forest Fire Detection: a Practical Use Case
\thanks{*These co-authors have been equally contributing the work.}
}

\author{
\IEEEauthorblockN{A. Belcher*}
\IEEEauthorblockA{\textit{Imperial College London} \\
\textit{UK}
}
\and
\IEEEauthorblockN{M. Esteva*}
\IEEEauthorblockA{\textit{Imperial College London} \\
\textit{UK}}
\and
\IEEEauthorblockN{A. Lam*}
\IEEEauthorblockA{\textit{Imperial College London} \\
\textit{UK}
}
\and
\IEEEauthorblockN{R. Ramadhani*}
\IEEEauthorblockA{\textit{Imperial College London} \\
\textit{UK}
}
\and
\IEEEauthorblockN{A. Rayhan*}
\IEEEauthorblockA{\textit{Imperial College London} \\
\textit{UK}
}
\and
\IEEEauthorblockN{W. Xu}
\IEEEauthorblockA{\textit{Imperial College London} \\
\textit{UK}
}
\and
\IEEEauthorblockN{D. Tuncer}
\IEEEauthorblockA{\textit{Ecole des Ponts ParisTech} \\
\textit{France}}
}

\maketitle

\begin{abstract}
The implementation of early warning mechanisms that can be used to detect forest fires in rural areas is essential to mitigate their deleterious effects, in particular by notifying local fire authorities to mount timely emergency responses. 6G-enabled Internet of Things (IoT) infrastructures are promising technological developments in that direction. However, in practice, the ability to detect forest fires in an effective way using distributed sensor nodes is challenging to achieve. In this short paper, we exemplify this challenge based on a case study that uses real data collected from the Low-Cost Internet of Things Sensor of Haze Air Quality Disasters in Communities in Thailand and Southeast Asia (SEA-HAZEMON) platform. The work is a preliminary step towards assessing the efficacy of a real-life fire detection system based on distributed sensor nodes. More generally, the objective is to develop a set of practical guidelines for the design of a 6G-enabled IoT-based fire detection mechanism. 
\end{abstract}


\section{Introduction}
\label{sect:introduction}

Amidst escalating climate change impacts, fostering proactive adaptation measures is an urgent discourse within vulnerable rural communities worldwide. According to the United Nation Environment Programme \cite{unep22}, the prevalence of forest fires has become a recurrent menace. It is expected to rise by 30\% by 2050, exacerbated by anthropogenic activity and the unpredictability of more frequented climate events, such as higher temperatures, droughts, and stronger winds. 

Networking technologies can support the implementation of mechanisms that enable the early detection of forest fires. Previous work has for instance investigated the use of wireless sensor networks \cite{zhang09}, UAV-enabled edge computing solutions \cite{fouda22} and optical remote sensing technologies \cite{barmpoutis20}. 6G networks are envisioned to enable the connection of a massive number of Internet-of-Things (IoT) devices that can serve as a platform for the deployment of fire detection mechanisms \cite{verma20}. 

The Low-Cost Internet of Things Sensor for Haze Air Quality Disasters in Communities in Thailand and Southeast Asia (SEA-HAZEMON) initiative \cite{seahazemon} is an IoT platform that provides real-time environmental data in the Southeast Asian region. It employs a Long-Range Wide-Area Networking (LoRaWAN) infrastructure that contains end-node sensors tasked with monitoring air quality and gateways to forward data for storage and analytics. A main objective of SEA-HAZEMON is to mitigate the deleterious effects of forest fires by distributing low-cost air quality sensors across rural regions that can be used to support early warning systems for wildfires \cite{lertsinsrubtavee22} by notifying local fire authorities to mount timely emergency responses. 

Detecting forest fires in an effective way using distributed sensor nodes is, however, a challenging process in practice. The placement of sensor nodes can be constrained by geographical factors, \textit{e.g.,} landscape, proximity to the built environment. These constraints not only condition the deployment, maintenance, and operation cost of sensors, they also have an incidence on the efficacy of the sensor platform in terms of detecting events. For instance, while Storey \textit{et al.} \cite{storey22} suggest that particle pollution that is representative of the occurrence of a fire can be detected up to 50 km away, especially for fires with a surface of more than 1,000 ha, recent results obtained with the SEA-HAZEMON platform \cite{lertsinsrubtavee23} show that the detection range is only up to 5 km in distance. In addition, in the case of sensors located in urban or semi-urban environments, the variations of air particles that they monitor over time can be attributed to different types of human-induced activities, which can make the detection of a fire event a complex task.     

In this short paper, we exemplify this complexity in the practical use case of the fire detection system deployed through the SEA-HAZEMON platform in the Tak province of Thailand. The Tak province is located in the north of Thailand. It hosts several national parks and wildlife sanctuaries. The SEA-HAZEMON platform has a sensor node deployed in the town of Tak. The sensor monitors air quality through particulate matter ($PM_{x}$) measurements and is attached to a wind monitoring probe. We use a data-driven methodology that cross-analyzes the evolution of $PM_{2.5}$ (the most used proxy to characterize forest fires), wind velocity and directions and the occurrence of fires over 2022 to illustrate the complexity associated with exploiting sensor node monitored data to develop an efficient fire detection mechanism. The work is a preliminary step towards assessing the performance of the SEA-HAZEMON fire detection system. More generally, the objective of the work is to develop a set of practical guidelines for designing a 6G-enabled IoT-based fire detection mechanism. 

The remainder of the paper is organized as follows. Section \ref{sect:methodology} describes the data-based methodology developed in this work. Section \ref{sect:airQuality} investigates the evolution of air quality as monitored by the sensor node in Tak in 2022. Section \ref{sect:fireEvents} discusses the parameters that affect fire detection. Section \ref{sect:conclusions} provides some concluding remarks and pointers to our future work. 

\section{Data-driven methodology}
\label{sect:methodology}

To evaluate the efficacy of the SEA-HAZEMON platform in identifying and determining forest fires, we focus on the relationships observed in 2022 between air quality and wind measurements monitored by sensors, and fire hotspot information extracted for the same year from the Fire Information for Resource Management System (FIRMS) database \cite{firms} released by the NASA. In this work, we focus on data collected in Thailand. Some of the sensor nodes in Thailand are attached to wind monitoring probes, which is essential to understand the transfer of particulate matter ($PM_{x}$) from the location of the event to the detecting sensor when a fire occurs. 

\begin{figure}
	\centering
    	 \includegraphics[trim={10cm 1.5cm 10cm 4cm}, clip, scale=0.4]{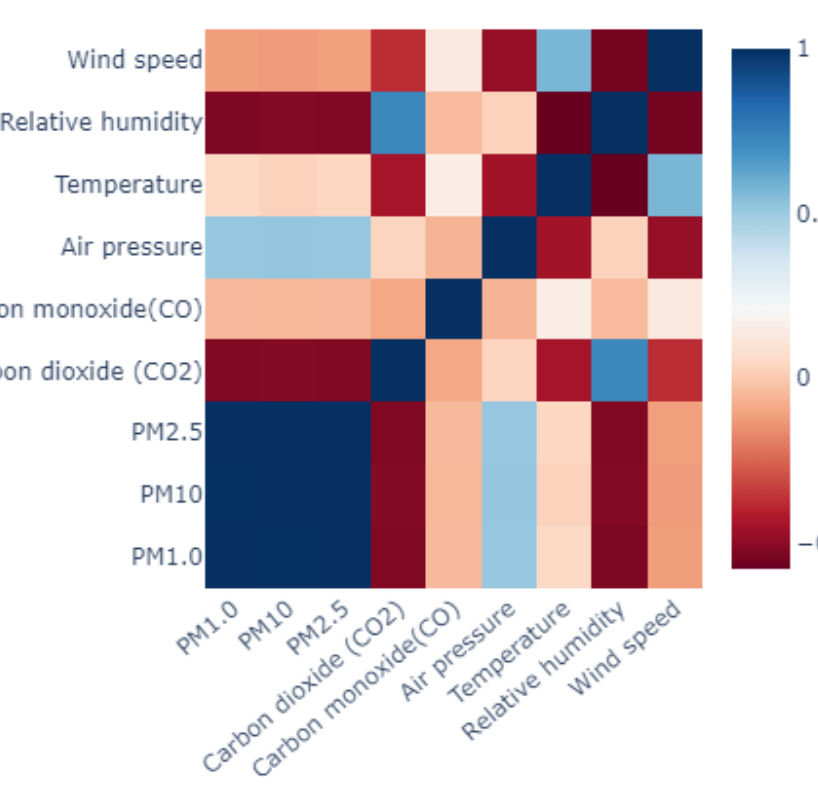}
	  \caption{Correlation matrix of pollutant levels and atmospheric parameters - Case of Moo5 sensor in the Tak province.}
	  \label{fig:corrMat}
\end{figure}

More specifically, we investigate two aspects. We first analyze the spatiotemporal variation of air particules over a year to understand how sensors capture different variations of $PM_{x}$ and $CO$ levels. We compare the measurements recorded by sensor node Moo5 in Tak with those collected by sensors located in six other provinces, \textit{i.e.,} Phetchabun, Chiang Rai, Prachuap Khiri Khan, Phayao, Lamphun, and Bangkok Metropolis. We then investigated the potential travel medium of $PM_{2.5}$ between the forest fire event and the Moo5 sensor node. 

The fire hotspot data for the Tak location is obtained from the FIRMS database \cite{firms} that uses the Moderate Resolution Imaging Spectroradiometer (MODIS) \cite{giglio21}. The primary attributes extracted for the analysis include the geographic coordinates (latitude and longitude) of the fire, the timing of imagery acquisition (Date-Time), and the Fire Radiative Power (FRP). FRP serves as a measure of the rate of emitted energy during a fire disturbance and is used as a proxy for the intensity of the fire.

Data collection and processing procedures are described in detail in \cite{belcher24}.

\begin{figure}
	\centering
    	 \includegraphics[trim=1cm 0.5cm 1cm 1cm, scale=0.45]{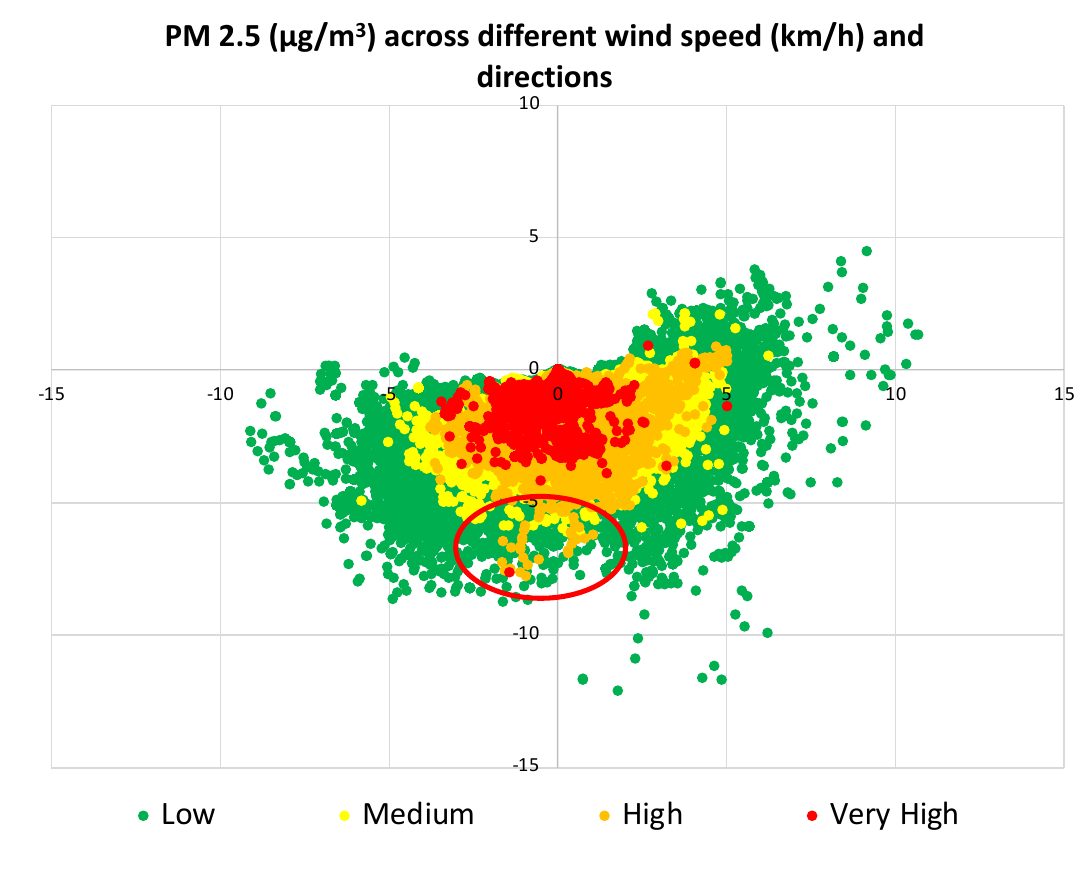}
	  \caption{Wind rose - Case of Moo5 sensor in the Tak province.}
	  \label{fig:windDirection}
\end{figure}

\begin{figure}
	\centering
    	 \includegraphics[trim=1cm 1cm 1cm 1cm, scale=0.25]{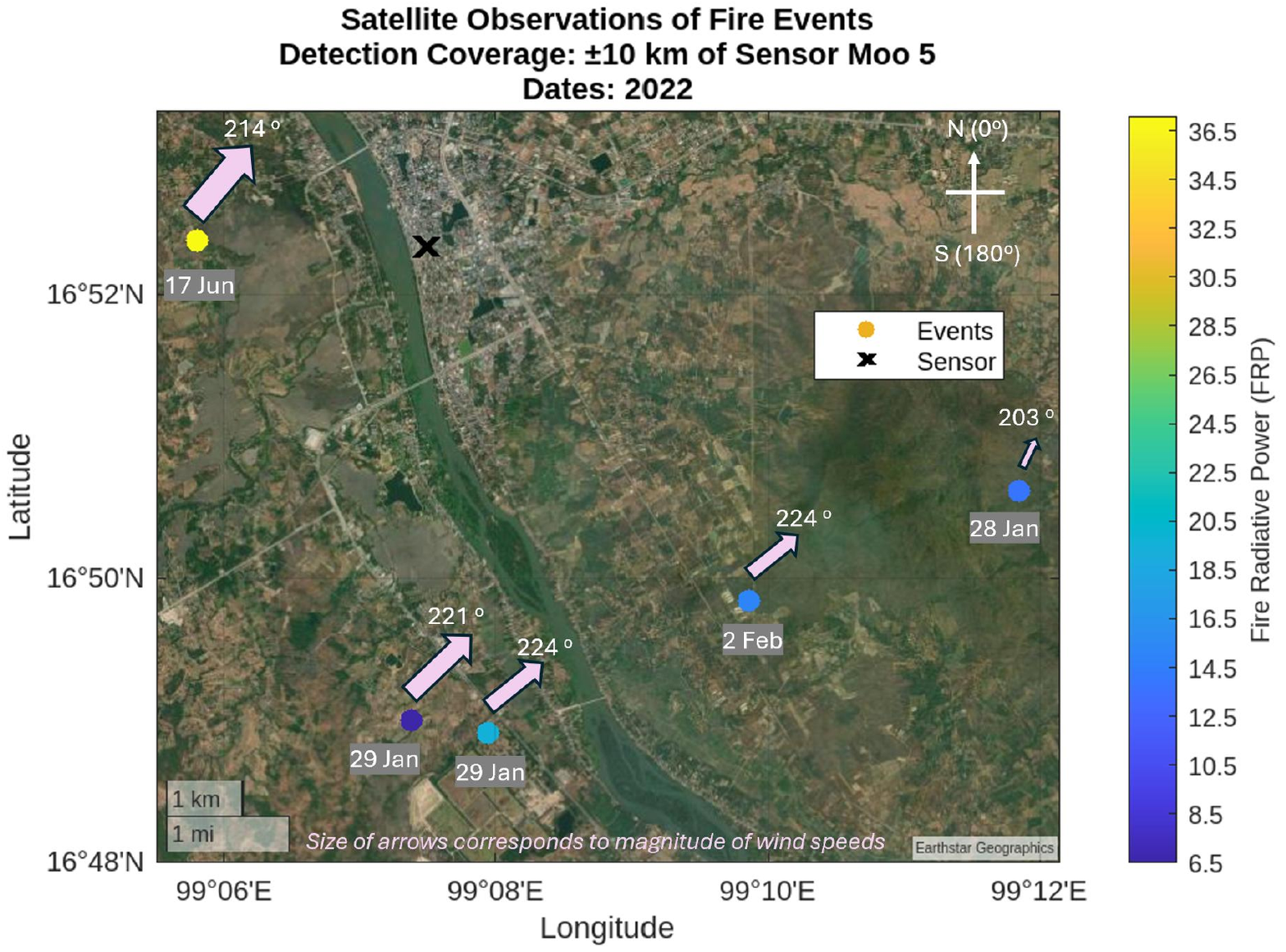}
	  \caption{Fire events mapped in relation to the Moo5 sensor node location.}
	  \label{fig:fireEvents}
\end{figure}

\section{Air quality evolution}
\label{sect:airQuality}

\begin{figure*}[h!]
	\centering
    	 \includegraphics[trim=1cm 10cm 1cm 1cm, scale=1]{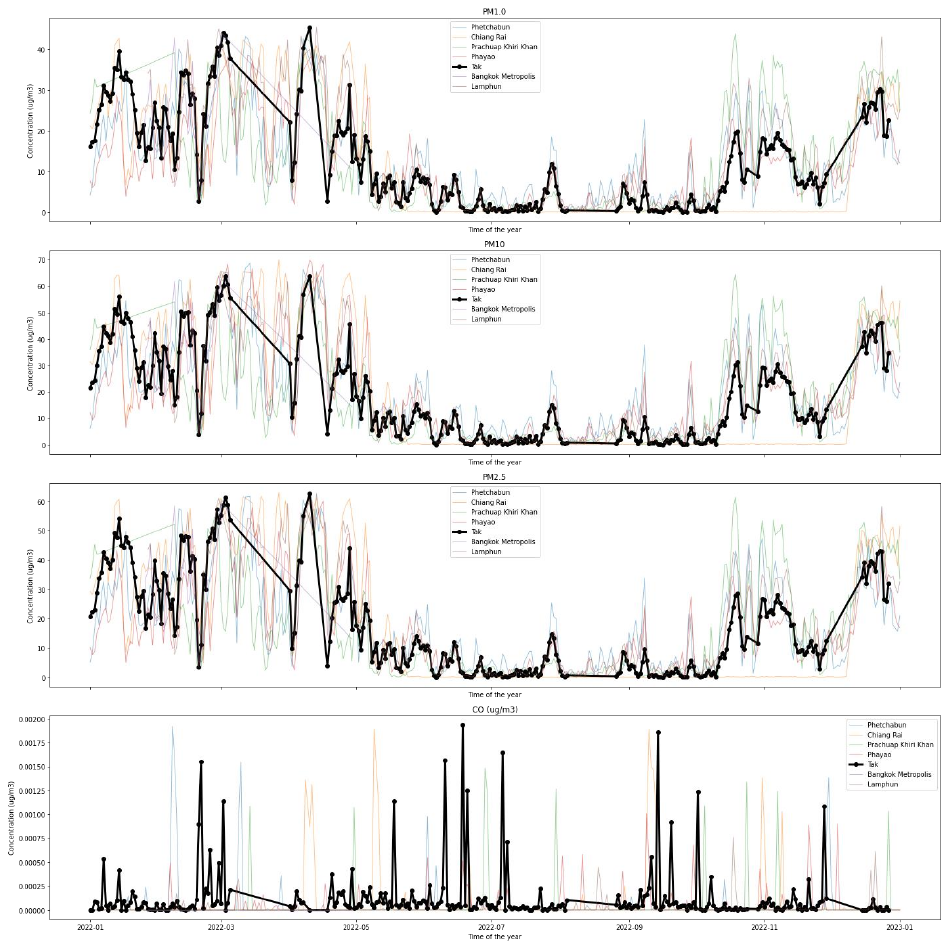}
	  \caption{Spatiotemporal variation of $PM_{x}$ and CO emission levels across different provinces in Thailand.}
	  \label{fig:annualPMLevel}
\end{figure*}

To understand how sensors react in practice to air quality variations, we analyze the evolution of $PM_{x}$ and CO levels across the different months of 2022 in Tak and the six other provinces. Figure \ref{fig:annualPMLevel} shows the spatiotemporal variations of $PM_{1.0}$, $PM_{10}$, $PM_{2.5}$ and $CO$ using a heatmap representation as recommended by Liu \textit{et al.} \cite{liu23}. As can be observed, $PM_{x}$ levels vary in a similar fashion in each province. High concentrations of $PM_{x}$ occur mainly in the January-April time period, \textit{ that is,} during the smoke season in all provinces. The observations are in line with previous work that shows that air pollutants in the upper north of Thailand have clear seasonal patterns \cite{sukkhum18}. Thus, these variations can be imputed to cyclical farming residue burning and arid weather conditions \cite{chansuebsri22}. In contrast to $PM_{x}$, $CO$ concentrations peak throughout the year. The data show that human-induced activities have an impact on the evolution of air quality patterns.  

To more specifically evaluate the relationships between different pollutant levels and atmospheric parameters, we calculate the Pearson's correlation heatmap for the case of the Moo5 sensor located in the Tak region, as shown in Figure \ref{fig:corrMat}. Strong negative correlation between $PM_{x}$ and relative humidity in Tak is consistent with the previous research effort conducted by Srithian \textit{et al.} \cite{sirithian22} in northern Thailand. High humidity enhances the deposition of $PM_{x}$, reducing the number of suspended particles. Similarly, wind speed tends to negatively correlate with pollutant levels. Although characterized by a moderate degree of correlation, high wind speed is expected to disperse pollutants in the air. However, the validity of this observation can vary due to several factors: proximity to pollution sources, topography, and atmospheric conditions. We plot the relationship between wind and $PM_{2.5}$ in Figure \ref{fig:windDirection} using the air quality boundary criteria set by the Thai Government \cite{unep18}, \textit{i.e.,} from low to very high concentration of $PM_{2.5}$. It can be observed that poorer air quality tends to be recorded when wind speeds are lower (red dots concentrated in the middle of the figure). We do, however, notice a few outliers (red circle) that we further investigate in the next section. 

\begin{figure}
	\centering
    	 \includegraphics[trim=1cm 0cm 1cm 1cm, scale=0.45]{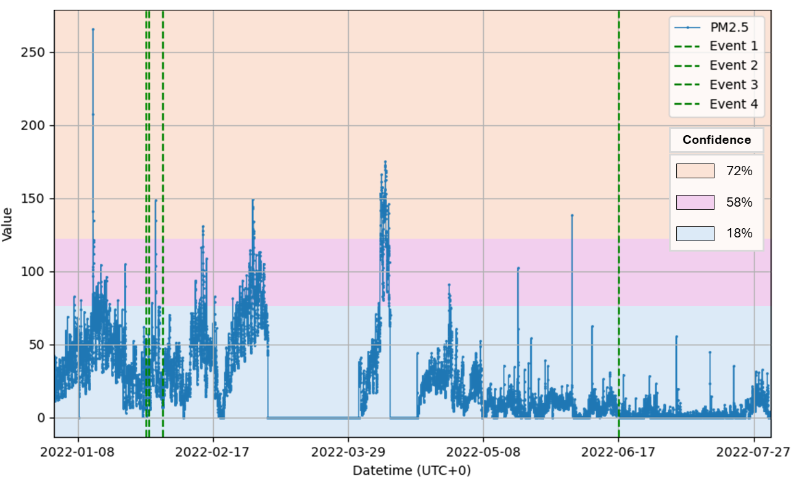}
	  \caption{Evolution of $PM_{2.5}$ concentration level and identified fire events.}
	  \label{fig:trueFires}
\end{figure}

\section{Fire event detection}
\label{sect:fireEvents}

We also investigated the ability of the Moo5 sensor to detect fire events. In this section, we focus on the concentration of $PM_{2.5}$ as the most widely used proxy to characterize forest fires. Figure \ref{fig:fireEvents} overlays on a satellite map showing the type of areas (vegetated, built), the five fire events recorded in 2022 (colored round shapes) in the vicinity of the Moo5 sensor (marked as a black cross). Fires are represented with their FRP and are accompanied by the wind direction at the time of occurrence (pink arrows). Considering both the distance of each fire event from the sensor node and the prevailing wind speed, a theoretical detection time can be estimated to help anticipate the temporal dynamics of the $PM_{2.5}$ dispersion. The satellite base map enables us to differentiate between forest fires and other types of fires based on whether the marker is on vegetated areas or man-made surfaces, such as roads. This distinction is crucial to accurately confirm the source of emissions from green landscapes.

\begin{figure}
	\centering
    	 \includegraphics[trim=1cm 0cm 1cm 0cm, scale=0.45]{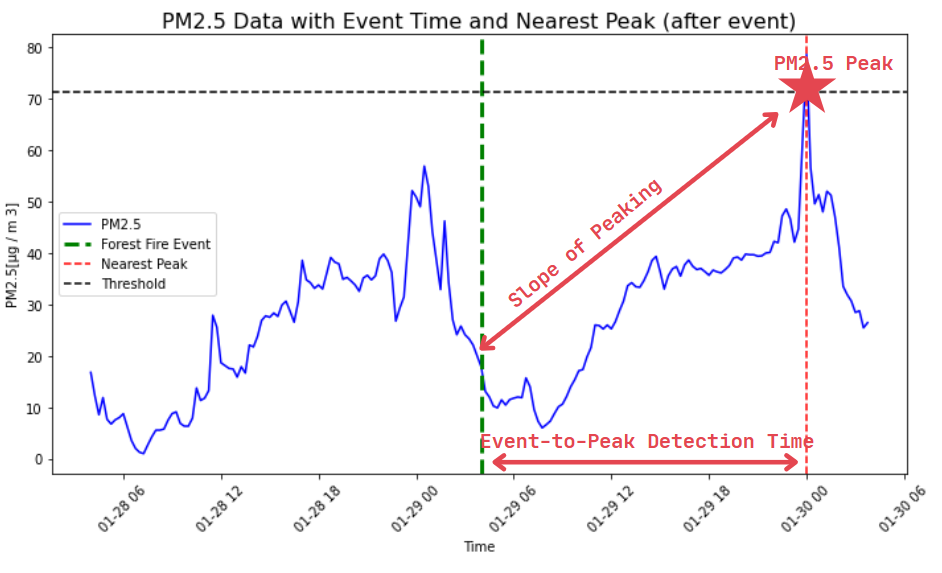}
	  \caption{Annotated time series of the $PM_{2.5}$ level during the double forest fires recorded in the Tak region on the 29th January, 2022.}
	  \label{fig:eventDetection}
\end{figure}

\begin{figure}
	\centering
    	 \includegraphics[trim=1cm 0cm 1cm 0cm, scale=0.45]{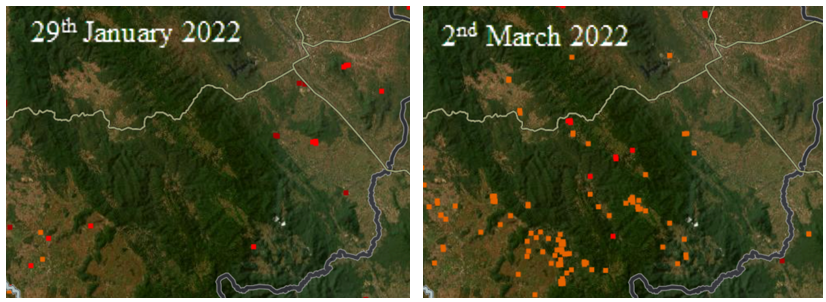}
	  \caption{Maps of fire event in the Tak region in 2022.}
	  \label{fig:fireMaps}
\end{figure}

\subsection{System capability}

SEA-HAZEMON uses an algorithm to detect forest fires that is based on setting thresholds of $PM_{2.5}$ and $CO$ concentrations. It employs a non-parametric supervised machine learning approach, trained using forest fire and emissions data from 2021. An event is marked as \textit{fire} when the $PM_{2.5}$ concentration exceeds the $71.22$ $\mu$g/$m^{3}$ threshold, with a reported confidence of 58\%.  

To verify the capability of the warning system, we overlay the 2022 year-long time series of $PM_{2.5}$ measurements with the identified fire events located within a 10 km radial distance from the sensor. Our objective is to assess whether the recorded fire event times correlate with the peak concentration of $PM_{2.5}$. The results are shown in Figure \ref{fig:trueFires}. As can be observed, not all peaks of $PM_{2.5}$ concentration levels are fire-related. Out of the five fire events in 2022, only the January 29th double fire event was followed by a peak of $PM_{2.5}$ exceeding SEA-HAZEMON's $71.22$ $\mu$g/$m^{3}$ threshold. Several peaks falling into in the pink and yellow zones in Figure \ref{fig:trueFires} could thus have sent false-positive fire alerts.

\subsection{Assessing sensor efficacy}

We then focus on the fire event that took place on January 29th 2022. We extract a 24-hour data window before and after the recorded time of the event (\textit{i.e.,} from January 28 to January 30), with $PM_{2.5}$ data aggregated into 15-minute intervals. We calculate the three following parameters: 1) the $PM_{2.5}$ level at the time of the event, 2) the subsequent peak concentrations of $PM_{2.5}$, and 3) the time taken for the $PM_{2.5}$ levels to reach the peak, which is indicative of the efficacy of SEA-HAZEMON's detection platform. Given that the theoretical detection time would be on the order of wind speed, wind speed and direction are also quantitatively determined to compare the time differences previously computed. The result is shown in Figure \ref{fig:eventDetection}. Although the average wind speed was 4.14 km / h, there was no discernible increase in the concentration level of $PM_{2.5}$ in the immediate following time window. Concentrations of $PM_{2.5}$ exceeded the threshold 20 hours later only, following a trend similar to the one recorded in the previous 24 hours (28th January). This suggests that the $PM_{2.5}$ peak was not necessarily caused by the fire event.

As shown in Figure \ref{fig:fireEvents}, the sensor cannot detect other fire events with different FRP values due to their location and wind direction. For instance, while the fire event that occurred in June 26th had the largest FRP compared to the other events, only a negligible increase of $PM_{2.5}$ concentration was recorded by the Moo5 sensor ($<$10 $\mu$g/$m^{3}$). In addition, while the direction and velocity of the wind have generally an incidence on $PM_{2.5}$ levels, Figure \ref{fig:windDirection} depicts a cluster of a few data points ($PM_{2.5}$ value circled in red) for which high wind leads to poor air quality. This cluster of points was observed only once (28th Feb – 1st Mar). Looking at the fire information for this period, as depicted in Figure \ref{fig:fireMaps}, it can be seen that many fires were recorded in the south of the Moo5 sensor, from the direction of the prevailing wind. These are significantly different in terms of their location from the fires that occurred within 10 km of the Moo5 sensor and were not captured by the node. 

\section{Conclusions}
\label{sect:conclusions}

The analysis of the measurement data obtained from IoT-based SEA-HAZEMON platform illustrates seasonal variations and peaks of $PM_{2.5}$ concentration in the Tak region of Thailand. The overlay analysis of fire timestamps in a 15-minute aggregated time interval of observation for the values of $PM_{2.5}$ suggests that none of the $PM_{2.5}$ peaks exceed the SEA-HAZEMON's fire detection threshold used to signify true forest fires. In addition, the closest in time observed $PM_{2.5}$ peak for all events spans from 15 to 21 hours after the fire, whereas the theoretical detection time is driven by wind speed. Our preliminary results show that the deployment of sensors in a semi-urban / urban location can create false positives. 

In the next step, we will complement our analysis by taking into account the variation of $CO$ levels. We will also investigate other sensor deployment locations in Thailand where wind data measurements are available. These results will enable us to derive a set of recommendations for the implementation of sensor nodes and the associated technological requirements for a 6G-enabled IoT infrastructure. 

\section*{Acknowledgement}

We thank Dr Adisorn Lertsinsrubtavee at intERLab in Thailand for his help going through the SEA-HAZEMON platform.


\begin{thebibliography}{00}
\bibitem{unep22} ``Spreading like Wildfire: The Rising Threat of Extraordinary Landscape Fires,'' United Nation Environment Programme, 23 February 2022. [Online]. Available: \url{https://www.unep.org/resources/report/spreading-wildfire-rising-threat-extraordinary-landscape-fires} - Accessed 10-06-24.
\bibitem{zhang09} J. Zhang, W. Li, Z. Yin, S. Liu, and X. Guo, ``Forest fire detection system based on wireless sensor network," in proc. \emph{4th IEEE conference on industrial electronics and applications}, pp. 520-523, 2009.
\bibitem{fouda22} M. M. Fouda, S. Sakib, Z. M. Fadlullah, N. Nasser, and M. Guizani, ``A lightweight hierarchical AI model for UAV-enabled edge computing with forest-fire detection use-case," \emph{IEEE Network}, vol. 36, no. 6, pp. 38-45.
\bibitem{barmpoutis20} P. Barmpoutis, P. Papaioannou, K. Dimitropoulos, N. Grammalidis, ``A review on early forest fire detection systems using optical remote sensing," \emph{Sensors}, vol. 20, no. 22, pp. 6442, 2020.
\bibitem{verma20} S. Verma, S. Kaur, M. A. Khan, and P. S. Sehdev, ``Toward green communication in 6G-enabled massive Internet of Things," \emph{IEEE Internet of Things Journal}, vol. 8, no. 7, pp. 5408-5415, 2020.
\bibitem{seahazemon} Low-Cost IoT Sensor of the Haze Air Quality Disasters in Communities in Thailand and Southeast Asia (SEA-HAZEMON). [Online] - Available at: \url{https://hazemon.in.th/v24.02/map.html?} - Accessed 10-06-24. 
\bibitem{lertsinsrubtavee22} A. Lertsinsrubtavee, KG. Sarambage Jayarathna, P. Mekbungwan, T. Kanabkaew, S. Raksakietisak, ``SEA-HAZEMON: Active Haze Monitoring and Forest Fire Detection Platform," in proc. of \emph{the 17th ACM Asian Internet Engineering Conference}, Hiroshima Japan, p. 88–95, 2022. 
\bibitem{storey22} M.A. Storey, O.F. Price, ``Statistical modelling of air quality impacts from individual forest fires in New South Wales, Australia," \emph{Natural Hazards and Earth System Sciences} vol. 22, no. 12, pp. 4039–4062, 2022.
\bibitem{lertsinsrubtavee23} A. Lertsinsrubtavee, T. Kanabkaew, S. Raksakietisak, ``Detection of forest fires and pollutant plume dispersion using IoT air quality sensors," \emph{Environmental Pollution}, vol. 338, pp. 122701, 2023. 
\bibitem{firms} Fire Information for Resource Management System (FIRMS). [Online]. Available: \url{https://www.earthdata.nasa.gov/learn/find-data/near-real-time/firms}  - Accessed 10-06-24.
\bibitem{giglio21} L. Giglio, W. Schroeder, J.V. Hall, ``MODIS Collection 6 and Collection 6.1 Active Fire Product User's Guide", 2021. 
\bibitem{belcher24} A. Belcher \textit{et al.} ``Assessing the Efficacy of IoT-based Forest Fire Detection: a Practical Use
Case - Data processing procedure", Technical report. [Online] - Available at: \url{www.dropbox.com/scl/fi/0fqifmsvnwoez87g28dik/DataProcessingProcedure_Belcher-et-al_2024.pdf?rlkey=rmlgfyx7uv9vp2jdbmaehty6t&dl=0} - Accessed 10-06-24.
\bibitem{liu23} J. Liu, G. Wan, W. Liu, C. Li, S. Peng, Z. Xie, ``High-dimensional spatiotemporal visual analysis of the air quality in China," \emph{Scientific Reports}, vol. 13, no. 1, pp. 5462, 2023. 
\bibitem{chansuebsri22} S. Chansuebsri, P. Kraisitnitikul, W. Wiriya, S. Chantara, ``Fresh and aged PM2.5 and their ion composition in rural and urban atmospheres of Northern Thailand in relation to source identification," \emph{Chemosphere}, vol. 286, pp. 131803, 2022.
\bibitem{sirithian22} D. Sirithian, P. Thanatrakolsri, ``Relationships between Meteorological and Particulate Matter Concentrations (PM2.5 and PM10) during the Haze Period in Urban and Rural Areas, Northern Thailand," \emph{Air, Soil and Water Research}, vol. 15. pp. 11786221221117264, 2022.
\bibitem{unep18} United Nation Environment Programme, Pollution Control Department, Department of Health, ``Air Quality Assessments for Health and Environment Policies in Thailand," 2018. 
\bibitem{sukkhum18} S. Sukkhum, A. Lim, T. Ingviya, R. Saelim, ``Seasonal Patterns and Trends of Air Pollution in the Upper Northern Thailand from 2004 to 2018," \emph{Aerosol Air Qual. Res.}, 22, 210318.
\end{thebibliography}
\end{document}